# The Nature of the Controversy over Time-Symmetric Quantum Counterfactuals

Ruth E. Kastner
University of Maryland, College Park

ABSTRACT. It is proposed that the recent controversy over "time-symmetric quantum counterfactuals" (TSQCs), based on the Aharonov-Bergmann-Lebowitz Rule for measurements of pre- and post-selected systems, can be clarified by taking TSQCs to be counterfactuals with a specific type of compound antecedent. In that case, inconsistency proofs such as that of Sharp and Shanks (1993) are not applicable, and the main issue becomes not whether such statements are true, but whether they are nontrivial. The latter question is addressed and answered in the negative. Thus it is concluded that TSQCs, understood as counterfactuals with a compound antecedent, are true but only trivially so,
and provide no new contingent information about specific quantum systems (except in special cases already identified in the literature).

1. Introduction.

Time Symmetric Quantum Counterfactuals are claims about the probabilities of outcomes of counterfactual (not-actually-performed) measurements on "pre- and post-selected" systems: that is,on systems identified by two measurement results at two different times $t_a$ and $t_b$, instead of the usual single pre-selection result at a single time ta. The current controversy over Time Symmetric Quantum Counterfactuals (TSQC) has its roots in a famous paper by Aharonov, Bergmann, and Lebowitz (henceforth, "ABL") entitled "Time Symmetry in the Quantum Process of Measurement" (1964).

The key concept introduced by ABL is that of a "pre- and post-selected ensemble," i.e., an ensemble of systems selected in a time-symmetric way via a preselection and then a second, final post-selection. The central result of the paper is a time-symmetric expression for the probability of an outcome of a measurement performed at a time t between such pre- and postselection measurements, subsequently known as the "ABL rule."

The ABL rule is a straightforward consequence of standard quantum theory in the case of actually performed measurements at all three times. It gives the probability of outcome $q_j$ of a nondegenerate observable Q measured at a time t between pre- and post-selection in states |a> at time $t_a$ and |b> at time $t_b$, respectively: (For simplicity and with no loss of generality, we consider the case of zero Hamiltonian):

$$P(q_j | a,b) = \frac{|\langle b|q_j\rangle|^2 |\langle q_j|a\rangle|^2}{\sum_i |\langle b|q_i\rangle|^2 |\langle q_i|a\rangle|^2} \qquad (1)$$

(1) is essentially a time-symmetric generalization of the von Neumann Projection Postulate or "Process 1" (von Neumann 1995). It assumes that the density matrix of the system at the intermediate time t is a proper or "ignorance"—type mixture of possible eigenstates of Q.

In 1985, Albert, Aharonov, and D'Amato (AAD) wrote a paper entitled "Curious New Statistical Predictions of Quantum Mechanics" which began a program of using the ABL rule to derive various results. This paper made a seemingly innocuous but unexamined assumption about the applicability of the rule, namely that it could be interpreted as applying to measurements that 'might have been carried out' (1985, 5). However, this apparently natural and innocuous assumption opened up a 'Pandora's Box' of controversy as to what kinds of statements are valid to make about pre- and post-selected systems.

As noted above, the ABL rule was derived on the assumption that the outcome whose probability is being calculated corresponds to a measurement that was actually performed during the pre- and post-selection process. However, AAD presented the ABL rule this way:

> "Consider a quantum mechanical system whose Hamiltonian, for simplicity, we shall take to be zero. Suppose that this system is measured at time ti to be in the state |A = a> (where A represents some complete set of commuting observables of the system, and a represents some particular set of eigenvalues of those observables), and is measured at time tf(tf > ti) to be in the state |B = b>. What do these results imply about the results of measurements that might have been carried out within the interval (ti > t > tf) between them? It turns out that the probability (which was first written down in ABL (1964)) that a measurement of some complete set of observables C within that interval, if it were carried out, would find that C = cj is
>
> $$P(c_j) = \frac{|\langle A = a | C = c_j \rangle|^2 |\langle C = c_j | B = b \rangle|^2}{\sum_i |\langle A = a | C = c_i \rangle|^2 |\langle C = c_i | B = b \rangle|^2},$$
>
> and that formula entails, among other things, that P(a) = P(b) = 1. Consequently, these authors maintain that that such a system, within such an interval, must have definite, dispersion-free values of both A and B, whether or not A and B may happen to commute." (original italics, boldface added for emphasis) (1985, 5)[1]

Now, it must be pointed out that this portrayal of the ABL rule is not consistent with the original derivation and presentation of the rule by ABL. Firstly, the phrases which I have highlighted in boldface, "might have been carried out" followed by a subjunctive or counterfactual conditional statement of the ABL rule, is the original TSQC-type reading of the ABL rule (which one might liken to the key to Pandora's Box). This view of the ABL rule was a seemingly natural but as-yet-unjustified leap from the actual, somewhat restrictive assumptions behind the ABL derivation— i.e., an intervening measurement actually being performed and the process resulting in a pre- and post-selected ensemble depending in part on that particular measurement—to a much less constrained situation in which the pre- and post-selected ensemble was viewed as a well-defined entity in its own right which could be conceptually "held fixed" while the intervening measurement was regarded as variable.

---

[1] It should be noted that the case in which the counterfactual measurement is one which commutes with either the pre- or post-selection observable is a special one in which the corresponding TSQC fulfills a consistent history condition and can therefore be seen as uncontroversially valid. What is contested by critics is the general case (any observable considered at t).

Assuming that by the words "these authors maintain . . . ",AAD mean ABL, their statement is incorrect. In fact, as observed also by Sharp and Shanks (1993, 494, footnote 2) ABL never make any claim in their 1964 paper about a system having "definite, dispersion-free values" of noncommuting observables.

A few years later, Sharp and Shanks (1993) gave a proof intended to demonstrate that TSQCs give predictions inconsistent with quantum theory. Such proofs (Cohen (1995), Miller (1996)) have become part of the controversy and will not be addressed in detail in this paper, which aims to formulate the question in different (and, hopefully, illuminating) terms.[2]

The counterfactual usage of the ABL rule, as proposed by Lev Vaidman (cf. 1996–1999) and Ulrich Mohrhoff (cf. 2000, 2001), which I am calling a "Time Symmetric Quantum Counterfactual" (TSQC), consists (as in AAD) in applying the rule to cases in which Q was not actually measured

---

[2] For a detailed analysis and defense of the S&S proof, see Kastner (1999a). For Vaidman's response, see Vaidman (1999a).

at t. Vaidman's proposed wording of his TSQC is as follows (with minor changes in notation to match that used in this paper):

(1V) "If a measurement of an observable Q were performed at time t, then the probability for Q = $q_j$ would equal PABL(qj), provided that the results of measurements performed on the system at times $t_a$ and $t_b$ are fixed" (Vaidman 1999a, 6 (e-print version)).[3]

Mohrhoff's is as follows:

(1M) "If a measurement of observable Q were performed on system S between the (actual) preparation of the probability measure |a><a| at time $t_a$ and the (actual) observation of the property |b><b| at time tb, but no measurement is actually performed between $t_a$ and tb, then the measurement of Q would yield $q_j$ with probability PABL(qj|a,b)" (Mohrhoff 2001, 865).

     Vaidman (cf. 1996–1999) has used (1V) to obtain what he calls "elements of reality" for pre- and post-selected quantum systems. Mohrhoff (2000) has used the ABL rule in the form of (1M) to obtain what he terms "objective probabilities" for quantum systems. He sees these time-symmetric "objective probabilities" as the most informative and epistemologically complete kinds of probabilities attributable to quantum systems, in contrast to what he terms "subjective probabilities." The latter are generally time-asymmetric, and (as he defines them) pertain to situations which fail to take into account all facts (such as outcomes of future measurements). Now, presumably, in defining quantities such as "elements of reality" and "objective

---

[3] It should be noted that Vaidman considers the TSQC as applicable to a counterfactual measurement in the case when some different observable is actually measured at t. In contrast, Mohrhoff restricts his TSQC to the case when no measurement is actually performed at t (at least for the applicability of the TSQC for obtaining what he terms "objective probabilities"; cf. Mohrhoff 2001).

probability," Vaidman and Mohrhoff intend their TSQTs to have a highly nontrivial character: i.e., they should give meaningful contingent information about specific quantum systems.

2. The Controversy to Date.

The most recent installment of the controversy over TSQCs involved an exchange between myself (Kastner 2001) and Mohrhoff (2001). In Kastner (2001) I argued that (1M) fails to get around the proof by Sharp and Shanks (1993) (henceforth "S&S") which showed that predictions obtained from a counterfactual usage of the ABL rule conflict with quantum mechanics. The problem was that (1M) does nothing to actually "fix" the pre- and post-selection results in the way TSQC advocates require for evasion of the S&S proof (see section 4 below). In Kastner (2001) I suggested that perhaps what advocates of TSQCs really had in mind by talk of "fixing" the pre- and post-selected states of systems subject to TSQC claims was what was referred to therein as Statement (1'):

(1') "Consider system S having pre- and post-selection results a and b at times ta and $t_b$ when a measurement of observable Q was not performed. If a measurement of observable Q had been performed at time t, $t_a < t < t_b$ on S, and if S had the same pre- and post-selection outcomes as above, then outcome $q_j$ would have resulted with probability PABL($q_j$|a,b)."
I noted that (1') was essentially equivalent to a weaker version of (1M), called (2):
(2) "In the possible world in which observable Q is measured and system S yields outcomes a and b at times $t_a$ and $t_b$ respectively, the probability of obtaining result $q_j$ at time t is given by PABL($q_j$|a,b)."

In his response, Mohrhoff (2001) did not address statement (1'), but indicated that he saw no difference between the statements (1M) and (2). If (1') and (2) are equivalent, it appears a reasonable assumption to take (1') as the intended meaning of his TSQC. Taking a TSQC claim to be equivalent to statement (1') means understanding talk about "fixing" pre- and post-selection results (such as in Vaidman's version (1V)) as equivalent to the second, italicized "if"-clause or antecedent in (1') (since both Vaidman and Mohrhoff acknowledge that post-selection results can't actually be "fixed").[4]

It seems that what the TSQC really is, then, is not just a simple counterfactual but rather a type of "compound" counterfactual—that is, one with a double antecedent. For the immediate purpose of clarifying the controversy over the correctness of TSQCs, I shall take (1') as the intended meaning of TSQCs. I shall defer the question of whether the above statements differ (I think that (1') and (2) are equivalent, and that (1) differs from both of those), which will be addressed in Section 4.

Statement (1') is of course undeniably true: If I were to measure an observable that was not actually measured at t, and if the system under discussion were pre- and post-selected with the same results as in the actual world, then of course the ABL rule would apply to the probabilities of outcomes of the not-really-measured observable. I certainly do not disagree with this assertion, nor, I think, do any of the other critics of

---

[4] For example, Vaidman says "In the counterfactual world in which a different measurement was performed at time t, the state before t is invariably the same, but the state after time t is invariably different (if the observables measured in actual and counterfactual worlds have different eigenvalues.) Therefore, we cannot hold fixed the quantum state of the world in the future." (Vaidman 1999a, 5 (e-print version)). Mohrhoff, in his (2001, 867), says: "Obviously, Dr. X [the experimenter] might not have obtained the result |b><b| at the time $t_b$ [were the counterfactual intervening measurement performed]".

TSQCs.[5]

Apparently, then, we have found a statement of the TSQC which is true. The trouble is that the addition of the second antecedent makes it much too weak to support the kinds of claims being advanced by advocates of the TSQC, such as the claim of a nontrivial "objective probability" by Mohrhoff, or of "elements of reality" by Vaidman. Why should the second antecedent make such a difference to the strength of the counterfactual statement? As a starting point, consider an everyday counterfactual such as (leaving aside for the moment possible objections that everyday counterfactuals can have nothing in common with ostensibly more exotic TSQCs, which I will address in the next section):

(A) If there had been a raffle this Wednesday, then nobody would have won.

Now, claim (A) is quite a surprising claim, since ordinarily, if one holds a raffle, there is almost certainly some winner (if only a sympathy entry from the person donating the raffle item). So (A) is quite a strong and dramatic and surprising claim as it stands.
Now consider the following variant of (A):

(A') If there had been a raffle this Wednesday, and if nobody had entered, then nobody would have won.

It is obvious that the addition of the second, auxiliary antecedent, "if nobody had entered," weakens the claim so much as to make it completely vacuous. It is, of course, undeniably true, but only trivially so: all it does is to restate the raffle rules. That is, it gives no specific information about the event under consideration: namely, a particular hypothetical raffle possibly held on a specific date, with certain specific potential entrants. But what is (A') other than a version of (A) with certain background conditions taken as "fixed"? For, if no raffle takes place on Wednesday, then in the actual world, there are no entrants. If we hold this condition fixed, then we obtain a consequence that there are no winners in a counterfactual raffle.

---

[5] The reason we have been critics is because of the claims based on the proposed TSQCs; cf. Sections 1 and 4.

The triviality of (A') is best explained in terms of what is called "cotenability" in many theories of counterfactuals. I shall follow Horwich (1988, Chapter 10) in describing the concept of cotenability with respect to counterfactual statements. The first serious attempt to construct a theory of counterfactuals was by Nelson Goodman (1947). Goodman proposed that a counterfactual statement of the form

CF: "If it were the case that P, then it would be the case that Q," symbolized by

$$P \,\square\!\!\rightarrow Q, \qquad (2)$$

is true if and only if the antecedent P, together with certain background conditions S holding when P is false, nomologically entail Q; in symbols:

$$P \& S \rightarrow Q. \qquad (3)$$

This formulation successfully captures the idea that a counterfactual is true when the "stage is set" (S) for something (Q) to happen if P were true (which it is not).

However, the notorious problem is in delimiting the background conditions S. As Horwich notes, we can make any absurd claim Q by using S to create a false conjunction on the left hand side of (3). For example, suppose in the actual world it is not raining. Then if I consider that fact as background condition S and introduce the antecedent P=not-S="It

is raining", I can obtain the following absurd conclusion:

(4) "If it is raining and not raining, then pigs can fly."

Statement (4) is vacuously true, but true nonetheless. In order to solve this problem, Goodman found that facts allowed in S had to be "cotenable" with P, which meant that they could only be those whose truth would not be affected by the truth of P. in other words, they had to fulfill the requirement

$$P \,\square\!\!\rightarrow S, \qquad (5)$$

which unfortunately made Goodman's definition of counterfactuals circular. (Note that (5) would successfully eliminate (4), since one could not maintain that "If it were raining, then it would not be raining.")

Nevertheless, we can apply the cotenability concept to see why (A') is trivial. A counterfactual is considered true because the "stage is set" (S) so that one additional event P, if true, leads nomologically to the consequent Q. But if the stage setting (S) has any dependence on whether P occurs, then the left hand side of (3) could be false, and Q may not occur nomologically.

In symbols, if S is no longer true when P is true, the conjunction (S&P) becomes false. But we needed the truth of (S&P) in order for Q to follow nomologically. If (S&P) is false, then any Q whatsoever follows vacuously, rather than nomologically, as in example (4). If (S&P) might be false, then Q might not follow nomologically, so we can't assert that it "would", which is what the counterfactual does (see again statement CF, which asserts "If it were the case that P, then it would be the case that Q," rather than merely "If it were the case that P, then it might be the case that Q").

Now, back to statement (A'). The role of the auxiliary antecedent in

(A') is to get around the failure of cotenability between the antecedent (a raffle is held) and the background conditions S in place when the antecedent is false (nobody enters). Cotenability fails because we cannot assert (5) in this case:

$$P \,\square\!\!\rightarrow S$$

(where P is "a raffle is held" and S is "there are no entrants") because when raffles are held, people generally enter them. That is, the best we can do is to say "If a raffle were held, there might be no entrants," with S being a highly unlikely occurrence. Thus the introduction of the antecedent P affects the truth of those background conditions S, so those background conditions are not cotenable with P. The only way we can force Q to nomologically follow in the form (3) is to stipulate that those background conditions don't change upon introduction of P, despite the fact that they normally would. To do this we add an auxiliary antecedent stipulating the certainty of background condition(s) S which would not normally be certain. In making such a stipulation, we invoke a state of affairs that conflicts with the known processes of our world (such as: when raffles are held, people enter them; and when measurements are made at time t, outcomes at time $t_b$ generally don't occur with certainty but only with some probability dependent on the measurement outcome at time t).

    Note that there can be nontrivial compound counterfactuals, and it is not required that the main antecedent and the auxiliary antecedent guarantee each other. All that is required for cotenability is that the background conditions holding when the antecedent is false (in this case, no entrants) have no dependence on the truth of the main antecedent. In the specific case of (A'), whether or not there are entrants does depend on whether a raffle is held, and that is why cotenability fails and (A'), which stipulates background conditions that normally would not hold, becomes trivial.

The following is an example of a nontrivial compound counterfactual:[6]
(B): "If a raffle were held and if three people had entered, then with probability x one of them would have won."

In (B), the quantity x can take on different values and in so doing will give different information about how the raffle is run (i.e., different rules about the nature of the pool of tickets, purchased and unpurchased, from which the winning ticket is drawn). Note that the second antecedent in (B) is not the background condition S (no entrants) in place when the antecedent was false, so it does not have the same structure as (A') or (1'). The crucial point is that in (B), non-cotenable background conditions S (no entrants) are not being invoked in order to to obtain the truth of the consequent.

Here is another compound counterfactual that is nontrivial[7]: Suppose the only people in town this week, besides myself, are such that they buy raffle tickets but don't claim prizes (in other words, only the absentminded people are in town). Then one might make the following counterfactual claim:

(C): "If there had been a raffle, and if only I and the absent-minded people had entered, then I would have won the raffle."

Again, the above is not completely trivial because the second antecedent does not serve the purpose of circumventing noncotenability. The background condition holding when the antecedent is false (i.e., when no raffle is held) is that no one enters the raffle, and this not what the second antecedent asserts. But notice that what the second antecedent asserts is extremely unlikely (how plausible is it that everyone besides me who enters and wins a raffle fails to claim their prize?), and that the ordinarily surprising conclusion "I would have won the raffle" is therefore much less surprising. It is less surprising in proportion to the extent that I have "tampered" with the background conditions (i.e., tailored them to my desired outcome of winning the raffle).

---

[6] (B) was suggested by an anonymous referee.
[7] As suggested by another anonymous referee.

To sum up: if the TSQC (1') is understood as a counterfactual with an auxiliary antecedent whose function is to fix the statement of background conditions S holding when P is false, where S is not cotenable with the main antecedent, it is only trivially true in the same way that (A' is trivially true.

Therefore quantities obtained from TSQCs that appear surprising at first glance, such as probabilities of unity for outcomes of noncommuting observables, are laden with such a heavy burden of conditions that they cease to apply to systems under study in any sort of physically meaningful way (just as no one cares about the fact that, if nobody entered a raffle, then nobody would win).

3. Are TSQCs Immune to the Charge of Vacuity?

The argument thus far has been that a TSQC, understood as (1')—a counterfactual with explicitly "fixed" noncotenable background conditions—has the structure of statement (A'), which is obviously trivial. Thus it is an argument by analogy: the triviality of (A') underscores the triviality of (1'). (In fact it seems to this author that the triviality of (1') is already clearly evident, since the auxiliary antecedent is such a "big if.") The persuasiveness of this argument therefore depends on the strength of the analogy. TSQC advocates insist that ordinary counterfactuals are "classical," and maintain, on that basis, that TSQCs are immune from any analogies with ordinary counterfactuals. However, this claim will be challenged in what follows.

Here we apparently need to address the specific ontologies proposed by Vaidman and Mohrhoff, since they both claim that their TSQC proposals provide new insights into the nature of the quantum world. Therefore they obviously think TSQCs are nonvacuous. Thus our task boils down to deciding whether quantum systems and the physics describing them, according to TSQC proponents, makes them immune to the triviality of stating that if no one entered a counterfactual raffle, then no one would win.

For starters, then, let us first adopt Mohrhoff's viewpoint, which, as I understand it from his (2000) and (2001), is characterised by the following key beliefs:

a. No time index applies to unobserved quantum systems.
b. No intrinsic properties are possessed by unobserved quantum systems. The only time that a system can be said to "possess" a property is when a measurement with a definite outcome has occurred.
c. There is no "flow" of time at the microscopic level, either forward or backward.
d. There is no real difference between past, present, and future.
e. There is no causality.

It should be noted that (a) through (e) constitute a basic metaphysical position concerning events in time, and that Mohrhoff's claims about objective probabilities are secondary to these basic assumptions. Therefore Mohrhoff sees his usage of the ABL rule in deriving what he calls "objective probabilities" as justified by these assumptions.
The task now becomes to understand why, or if, beliefs (a) through (e) should make Statement (1′) any less vacuous than (A′). In other words, since everyone surely agrees that the statement concerning a "counterfactual raffle" with fixed background conditions, (A′), is vacuous, we need to see whether beliefs (a) through (e) serve to make statement (1′) immune from the same kind of vacuity. (The latter is essentially what Mohrhoff has argued in his defense of his usage of the ABL rule, which is why I feel compelled to address it here). So let us catalog the differences (according to Mohrhoff's ontological assumptions):

It is generally assumed that a raffle consists of physical systems that always have observable properties (i.e., possessed properties, as defined according to (b)). Two considerations arise: (i) it need not be the case that the systems involved in the raffle are classical systems, and (ii) even if it did, does the fact that a raffle involves observable (possessed) properties, causality and time flow throughout the interval [ta, tb] have a bearing on whether (1′) is as vacuous as (A′)?

In support of (i), consider a quantum raffle. It goes like this: at $t_a$ (say, Monday), there are N possible entrants (each of which could be some sort of device rather than a person). (We require that N be nonzero, otherwise it makes no sense to consider any kind of raffle, whether actual or not. Thus the number N becomes a component of the raffle rules.) Each prospective entrant holds a quantum coin. If no raffle is held at time t (Wednesday), the coin remains in an unflipped ready state. If a raffle is held at time t, a signal is sent to each of the N prospective entrants which triggers a coin flip. For each coin flip that comes up heads, there is an entrant. (We need not concern ourselves with how a winner is chosen from the pool of entrants; our question concerns only whether or not there is a winner.) At time $t_b$ (say, Friday), the number of entrants M is recorded. Obviously, when M = 0, there is no winner.

Now, when there is no raffle, there are no coin flips, therefore none comes up heads, therefore M = 0 and there is no winner. Statement (A′) asserts the obvious: namely, if a counterfactual raffle were held and if none of the coin flips came up heads, there would be no winner.

I would argue that the counterfactual quantum raffle described is isomorphic, in every relevant sense, to the situation considered in a TSQC. (Note that I am not claiming that the quantum raffle is the same procedure as in a TSQC; obviously it is not. All I am claiming is that the general form of the claim corresponds in every relevant sense to the TSQC.) First, regarding the quantum raffle: we have an empirical fact at $t_a$ (Monday): the number N of prospective entrants. Prior to $t_b$ (Friday), there is no fact of the matter as to possessed properties of the quantum coins (ready, heads or tails), since they are not being measured (according to Mohrhoff's ontology). Only at time $t_b$ do we measure the coins and find out how many (M) are in heads states. We then note that M=0, and conclude that, had a raffle been performed and if the same outcomes had obtained at $t_a$ and tb, then no one would have won the raffle.

In the TSQC case, we have an empirical fact at ta: the outcome of the pre-selection measurement. Prior to tb, there is no fact of the matter as to what properties the system has, since it is not being measured (again, according to Mohrhoff's ontology). Only at time $t_b$ do we measure the system and find out which eigenvalue of the post-selection observable obtains at that time. We are then in a position to input the outcomes observed at $t_a$ and $t_b$ into the ABL rule and conclude that, had a measurement of Q been performed at t, and if the same outcomes had obtained at $t_a$ and bb, then the probabilities of eigenvalues of Q would have been as given by the ABL rule.

One might find a difference in that the output of the TSQC applies to an outcome at time t (prior to tb) while the outcome of the raffle seems to apply to time $t_b$ (Friday). Against this, we reply that, assuming time symmetry and/or a lack of time index in either case—remember, this is a quantum raffle with a time-symmetric, antirealist ontology—that the raffle outcome can be seen as applying at time t, just as in the TSQC. Therefore those holding the raffle might not discover the bad news until Friday, but one can apply time-symmmetry to argue that "in fact" there was no winner prior to Friday (or, to put it slightly differently, that there was fated to be no winner prior to Friday).

One could also, of course, point out that the quantum raffle is a slightly different kind of experimental procedure than the usual situation considered in a TSQC. But again, this difference is superficial. To see this, let's fill out the details in a possible quantum raffle. Assume a three dimensional Hilbert space for which the "ready," "heads," and "tails" states form a basis. Then a raffle taking place at t corresponds to a unitary evolution of the ready state to a state which is an equal superposition of the heads and tails states, call it the "flipped" state ( $1/\sqrt{2}$ [|heads> + |tails>]). At tb, a measurement of an observable with outcomes [heads] or [notheads] is performed.

So the raffle differs from the usual TSQT in that there is a unitary evolution between t and $t_b$ if the raffle is held; but since such an evolution is fully time symmetric, the difference in no way disqualifies the example as a fair analogy.

But, considering point (ii), suppose there really is no fully time symmetric, nonclassical raffle? Suppose there must always be some "classical" component, whether a possessed property, a temporal direction, a causal influence, involved in processes leading to statements such as (A')? So what? While both Mohrhoff and Vaidman have insisted that certain "behind-the-scenes" features of quantum systems (i.e., questions of how it happens that a system ends up with one outcome or another at times $t_a$ or tb) are what immunize TSQCs from comparisons with everyday counterfactuals, no reason has been given for thinking that such behind-the-scenes features have any bearing on the legitimacy of the counterfactual claims under consideration. On the contrary, I show below that Mohrhoff's own definitions of his proposed applications of the ABL rule imply that behind-the-scenes features precisely fail to immunize TSQCs from comparison with everyday counterfactuals.

In his (2001), Mohrhoff invokes behind-the-scenes considerations as crucial to the validity of his TSQC. That is, he (and Vaidman, as noted above) rejects arguments against his TSQC if they seem to be based on everyday, "classical" counterfactuals (such as my statements A and A') which, he assumes, invariably involve systems that always possess properties during the time interval [ta, tb]. Specifically, he states that TSQCs differ fundamentally from ordinary "classical" counterfactuals (such as A) because the latter involve systems with determinate properties and the former involves systems with indeterminate properties unless measured: "While a classical counterfactual assumes that something obtains whereas in reality something else obtains, a quantum counterfactual assumes that something obtains where in reality nothing obtains" (Mohrhoff 2001, note 23).

But by this definition, Mohrhoff's own "subjective" counterfactual usage of the ABL rule would constitute a "classical counterfactual" which would therefore be disqualified from comparison with his TSQC: "In principle, both rules [Born and ABL] have an objective as well as a subjective application. If Q is actually measured, both rules assign probabilities that are subjective inasmuch as they are based on probability measures that fail to take account of at least one relevant fact—the result of the measurement of Q" (Mohrhoff 2001, 865).

In the case of what Mohrhoff terms the "subjective" counterfactual application of the ABL rule, in reality "something else obtains" at t. Now, the "subjective" counterfactual application of the ABL rule is clearly some kind of TSQC (though not yielding what Mohrhoff would term "objective probabilities"). Therefore he cannot coherently disallow (A') as a valid analogy with TSQCs based on the fact that the former might involve systems with determinate behind-the-scenes features.

In any case, it must be reiterated (recall point (i) above) that it is perfectly possible that an "ordinary" counterfactual statement such as A' can always be replaced by a suitably "indeterminate" version (such as a quantum raffle) and that the resulting statement is clearly just as vacuous. But (ignoring for the moment the subjective ABL counterfactual claim of Mohrhoff which shows (i) to be unnecessary anyway), even if (i) should prove difficult to fulfill, anyone with a universally antirealist ontology can do this for ordinary classical processes merely by asserting *esse est percipi*. Then (A') involves indeterminate properties just as much as (1') does. But I doubt that even Bishop Berkeley would regard Statement (A') as nonvacuous.

The fact that, no matter what one does with one's ontological assumptions, (A') remain just as trivial, should be taken as a good indication that varying one's ontological assumptions are not sufficient to rescue (1') (which has the same form as (A')) from vacuity. What about a possible objection that the thing that saves (1') from vacuity is precisely the ABL rule (as opposed to some other kind of rule, like that of the raffle)? That would mean that counterfactual statements with a compound antecedent, such as (A'), are indeed trivial unless the underlying rule is the ABL rule. But against this, it is clearly the form of the statement, not the content, which makes it vacuous. To put it differently, the computational rule is not what is at issue; rather it is the contingent, empirical numbers (arising from assumed background conditions) input in the rule that are at issue. Therefore the precise nature of the specific computational rule has no bearing on whether the statement is vacuous.
Thus, denying all properties, time, and causality between measurements does nothing to emeliorate the vacuity of (1'), because its vacuity stems only from the necessity to add an extra condition, and not from any assumed macroscopic, classical, determinate attributes. The extra condition is required simply because there are facts that require "fixing" for the consequent to follow, but which are physically not fixed (the latter fact being acknowledged by proponents of TSQCs). This has nothing to do with whether or not one subscribes to the beliefs (a) through (e); it is simply the violation of cotenability between those facts and the antecedent (Mohrhoff, in 2001, has denied a cotenability problem for TSQCs, but his argument is flawed; a detailed refutation is presented below in Section 5).

It should be noted that Mohrhoff's tenseless view of facts—i.e., that a statement such as "X is true at time tb" should be seen as holding at all other times—fails to accomplish the kind of counterfactual fixing he seeks. This is because, if we are going to consider a counterfactual event at t— an event that might have occurred but didn't—then, to be consistent with physical law, we also have to consider possible outcomes at either $t_a$ or tb other than the actual ones, that might have occurred but didn't. Mohrhoff acknowledges these other possible outcomes but calls them "irrelevant," which can only be justified if his TSQC contains the second antecedent appearing in (1'), which explicitly instructs us to disregard them. But the second antecedent removes the need for any arguments that facts are untensed and therefore fixed, since said facts—whether tensed or untensed— are being "fixed" by the additional antecedent, which throws out the unwanted possible outcomes, in either case.

The present author is currently agnostic regarding tensed vs. untensed views of facts, and wishes merely to point out that the untensed view does nothing to accomplish the goal of obtaining counterfactually fixed events at times $t_a$ and $t_b$ in the absence of the second antecedent (without violating quantum theory—in which case the ABL rule would not hold anyway, since it is nothing more than a deductive consequence of quantum theory).

We turn briefly now to Vaidman's ontology, which differs from Mohrhoff's in that it assumes a bi-directional causal flow: one in the reversed time direction originating from the measurement at tb, along with the usual "retarded" causal flow originating at ta. But such metaphysical precepts concern what happens "behind the scenes," and I have already argued that whatever goes on behind the scenes has no bearing on the vacuity of the counterfactual claim. Whether or not there is reversed causal flow from tb, the fact remains that the post-selection result at $t_b$ is not actually physically fixed (as Vaidman readily admits; see footnote 4), and this makes Vaidman's TSQC also isomorphic to statements (1') and (A').

Thus, if TSQCs are properly understood as Statement (1')—and I think they clearly are, as argued above—then they are completely vacuous. All they do is to restate the ABL rule, providing no contingent information about the specific systems under study. In the same way, statement (A') tells us nothing substantive about the nature of the specific people, devices, or raffle-holding entity, in place during the time interval in question [ta, tb] (Monday through Friday), but merely restates the raffle rules.

4. Are Statements (1) and (1') Different?

Recall that (1) has two variants: Mohrhoff's version, which I am calling (1M), and Vaidman's version, which I am calling (1V) (although Vaidman's TSQC is really closer in wording to (1')). Let us first consider (1M). Statement (1M) fails to correctly convey the meaning of the TSQC—if the intended TSQC is truly (1')—because it contains only a single antecedent, and lacks any statement of the necessary additional condition (the auxiliary antecedent) required for the validity of the claim. It merely restates (i) that the measurement at t is not performed in the actual world (redundant since we already know that the statement is counterfactual) and (ii) the pre- and post-selectionmresults occurring in the actual world, which we also already know. Thus (1M) is completely equivalent to the single-antecedent counterfactual:

(3) If I had measured Q at t, then the probability of outcome qk would be as given by the ABL rule.

(3) is a stronger claim than (1'), in the way that (A) is a stronger claim than (A'). That is, (3) and (A) are highly nontrivial (but generally false) counterfactual claims. Both of these omit the auxiliary antecedent condition required for the truth of the claim (but which also makes the claim trivial). (A) is obviously false; (3) is false as shown by the S&S proof. That is, in failing to explicitly "fix" the required outcome at tb, (3) permits the application of the Sharp and Shanks inconsistency proof which demonstrates that such claims (in general) contradict quantum mechanics. Therefore (3), which would unambiguously do the work desired by Vaidman and Mohrhoff (in giving us "surprising" probabilities and/or specific contingent objective probabilities) is (generally) false.

As for (1V), as noted above, if we understand the "fixing" requirement as equivalent to the additional condition referred to in the auxiliary counterfactual antecedent, then (1V) is simply equivalent to (1'). It is therefore
vacuous, meaning that quantities derived from it do not really apply to specific systems in the way in which it has been claimed. Criticisms of (1V) can be seen as directed to claims based on (1V), rather than to the vacuously true nature of (1V) itself. That is, either (i) TSQC proponents have essentially been proposing (3), which is false and therefore not applicable to quantum systems; or they have been proposing (1'), which is vacuous and therefore also yields no valid information about specific quantum systems. In either case, what continues to be invalid is the use to which

(1V) has been put in supporting claims such as that one can "Ascertain the Values of σx, σy, and σz of a Spin-1/2 Particle," (Vaidman, Aharonov, and Albert 1987) and other "surprising" effects (cf. Vaidman (1996b, 900–901). The 1987 title itself explicitly attributes values obtained from counterfactual usages of the ABL rule (i.e., values corresponding to observables that were not measured) to a single system.

In another example, according to Vaidman and Mohrhoff, the ABL rule can be applied both conventionally and counterfactually to a particular individual particle in the "Three-Box" experiment, yielding the "surprising" (Vaidman) or "objective" (Mohrhoff) result that the particular particle's probability of being located in box A is unity and its probability of being located in box B is unity, when only one or the other measurement (opening either box A or box B) is actually performed. Clearly that would be surprising and substantive information, but the ABL probability of unity corresponding to the measurement that was not performed—the counterfactual one—depends crucially on an auxiliary "if" clause fixing non-cotenable selection results, just as the surprising result of statement (A)—that no one would win a counterfactual raffle—depends crucially on fixing the non-cotenable background condition that nobody enters a raffle. In both cases, the "surprising" result ceases to be surprising or objectively applicable to the specific particle (raffle) once the background conditions are fixed "by hand" in this way. The above kinds of specific claims which depend upon TSQCs are thus seen to be invalid, even if the TSQC can escape the Sharp & Shanks proof in the formulation (1').

5. Cotenability Violation Not Addressed by Mohrhoff Argument.

In Section 2, I described how TSQCs are made trivial because they invoke an auxiliary antecedent to circumvent the fact that the necessary background conditions (both the pre- and post-selection outcomes) are not cotenable with the antecedent (a counterfactual measurement at t). Mohrhoff denies a problem with cotenability, and in his (2001) uses the impossibility of superluminal signaling in an EPR (Einstein-Podolsky-Rosen) experiment as an analogy.

 Before turning to that analogy, it should be noted that Mohrhoff's discussion misconstrues the meaning of cotenability as something much weaker than it is. He assumes that background conditions need only be consistent with a counterfactual event to satisfy cotenability, but this is not what cotenability means. To use the raffle as an example, the back ground condition of there being no raffle entrants ($M = 0$) is physically consistent with a counterfactual raffle, but not cotenable with it. It is consistent because it is possible (though highly unlikely) that no one would enter a raffle held on Wednesday. (I.e., as noted in Section 2, one can truthfully say "If a raffle were held, there might not be any entrants.") However, it is not cotenable because its (counterfactual, not actual) truth depends on whether or not a raffle is held; because when there are raffles, people enter them (or quantum devices flip coins). Cotenability is a much stronger requirement than mere consistency between background conditions and counterfactual events. As discussed above (refer to expression (5)), it requires that there be no counterfactual dependence of those background conditions on the truth of the antecedent.

 To return to Mohrhoff's analogy: in the famous example of a perfectly anticorrelated pair of spin 1/2 particles separated by a spacelike distance and measured by Alice and Bob, superluminal signaling is impossible because, as Mohrhoff says, "it is impossible for Bob not only to determine the spin component measured by Alice but also to find out whether or not any spin component is measured by Alice" (Mohrhoff 2001, note 21). This, of course, is because no initial information concerning spin orientation is available from the density matrix of each of the particles, which is just proportional to the identity.

Mohrhoff claims that an exactly analogous situation holds in a timelike sense as applied to a counterfactual measurement at time t, and that therefore there is no violation of cotenability between the background conditions holding at $t_a$ and $t_b$ and a counterfactual measurement at time t (loosely speaking, that those background conditions are unaffected by such a measurement). However, this analogy is flawed.
The precise analogy drawn by Mohrhoff is the following: instead of two anticorrelated particles separated by a spacelike distance, we have a single particle perfectly correlated with itself (in terms of spin direction) at two different times. Mohrhoff assumes that Alice might make a measurement at t and Bob makes the post-selection measurement at tb, and notes that it is impossible for him to tell, based on his measurement, whether Alice made a measurement at t. But this analogy is false, because it neglects the known pre-selection at time ta—which, unlike the uninformative initial state of the anticorrelated particles of the EPR case, contains very specific information about spin orientation. Given a known preselection, say |a>, if Bob measures the observable A at time $t_b$ and obtains a result other than *a*, then he knows with certainty that Alice made a measurement of a noncommuting observable at time t. Therefore Alice's measurement certainly "disturbs" the particle in the way that Mohrhoff denies throughout his (2001). Ironically, Mohrhoff makes exactly this point (in slightly different terms) in the previous note (Mohrhoff 2001, note 20).

The phenomenon that I discuss above and that Mohrhoff discusses in his note 20 is most dramatically observed in the case of photons encountering crossed polarizers. If one places a polarizer oriented along, say, direction x at point a and another oriented exactly opposite to it at point b, then no photons will pass the second polarizer. (Think of this as the actual experiment, with no measurement at t.) But if one were to place a polarizer in between the first two, at an oblique angle to both (a counterfactual measurement at t), photons would be able to pass the second polarizer. Thus anyone post-selecting photons through a measurement at point b can determine whether someone has inserted a third polarizer in between a and b. This is exactly a violation of cotenability: an intervening event renders previous background conditions uncertain.

6. Conclusion: TSQCs Are either False or Vacuous.

If one takes the TSQC as (1M), it is false because it fails to state the extra condition (auxiliary antecedent) needed for the consequent to follow nomologically (as stated, the consequent does not follow). The Sharp and Shanks proof (1993),
which assumes no auxiliary antecedent, can be understood as a proof of the falsity of (1M). Objections by both Mohrhoff and Vaidman to the S&S proof seem to turn on the issue of whether or not the proof has taken into account the "fixity" of pre- and post-selection results, so those objections can be seen as supporting (1') as an accurate statement of their TSQC. If one takes the TSQC as (1'), the Sharp and Shanks proof does not apply; the TSQC is then true, but only trivially so. It does nothing but restate the ABL rule, and cannot be considered as providing information about specific systems as claimed. Metaphysical precepts concerning the reality (or lack thereof) of quantum systems, time, or properties between measurements have no bearing whatsoever on these conclusions, which are based solely on the empirically observable conditions necessary for the consequent to follow from the stated antecedent(s).